\documentclass[pra,aps,twocolumn,showpacs,superscriptaddress]{revtex4-1}

\usepackage{epsfig}
\usepackage{mathrsfs}
\usepackage{amsmath}
\usepackage{float}
\usepackage{color}

\begin{document}

\title{Protocol for Direct Counterfactual Quantum Communication}

\author{Hatim Salih}
\email{salih.hatim@gmail.com}
\affiliation{The National Center for Mathematics and Physics, KACST, P.O.Box 6086, Riyadh 11442, Saudi Arabia}

\author{Zheng-Hong Li}
\affiliation{The National Center for Mathematics and Physics, KACST, P.O.Box 6086, Riyadh 11442, Saudi Arabia}
\affiliation{
	Institute for Quantum Science and Engineering (IQSE)
and Department of Physics and Astronomy, Texas A\&M University, College Station, Texas 77843-4242
}

\author{M. \surname{Al-Amri}}
\affiliation{The National Center for Mathematics and Physics, KACST, P.O.Box 6086, Riyadh 11442, Saudi Arabia}
\affiliation{
	Institute for Quantum Science and Engineering (IQSE)
and Department of Physics and Astronomy, Texas A\&M University, College Station, Texas 77843-4242
}

\author{M. \surname{Suhail Zubairy}}
\affiliation{The National Center for Mathematics and Physics, KACST, P.O.Box 6086, Riyadh 11442, Saudi Arabia}
\affiliation{
	Institute for Quantum Science and Engineering (IQSE)
and Department of Physics and Astronomy, Texas A\&M University, College Station, Texas 77843-4242
}

\date{\today}

\begin{abstract}

It has long been assumed in physics that for information to travel between two parties in empty space, ``Alice'' and ``Bob'', physical particles have to travel between them. Here, using the ``chained'' quantum Zeno effect, we show how, in the ideal asymptotic limit, information can be transferred between Alice and Bob without any physical particles traveling between them.

\end{abstract}

\pacs{03.67.Hk, 03.67.-a, 03.65.Ta, 03.67.Dd, 03.67.Ac}

\maketitle

\newcommand{\ds}{\displaystyle}
\newcommand{\dd}{\partial}
\newcommand{\be}{\begin{equation}}
\newcommand{\ee}{\end{equation}}
\newcommand{\beq}{\begin{eqnarray}}
\newcommand{\eeq}{\end{eqnarray}}
\newcommand{\dt}{\ds\frac{\dd}{\dd t}}
\newcommand{\dz}{\ds\frac{\dd}{\dd z}}
\newcommand{\D}{\ds\left(\frac{\dd}{\dd t} + c \frac{\dd}{\dd z}\right)}

\newcommand{\w}{\omega}
\newcommand{\W}{\Omega}
\newcommand{\g}{\gamma}
\newcommand{\G}{\Gamma}
\newcommand{\E}{\hat E}
\newcommand{\s}{\sigma}
\newcommand{\bra}{\langle}
\newcommand{\ket}{\rangle}


Quantum mechanics has enjoyed immense success since its inception about a century ago. Its conceptual foundation, however, is often a matter of intense debate. Furthermore, several novel phenomena are predicted and observed based on quantum mechanics that appear counterintuitive and are unexplainable in the classical domain. Whole new fields owe their existence to this body of knowledge. One such field is quantum communication.
In this paper we propose a surprising mode of communication whereby no physical particles travel between sender and receiver.

In 1970, the idea of ``quantum money'' \cite{Qmon}- money that cannot be forged - came to light, effectively kick-starting the field of quantum information. The idea, too advanced for its time, rested on the conjecture that quantum states cannot be faithfully copied, which was later proved as the no-cloning theorem \cite{clontheo}. Moreover, the mere act of measurement of an unknown quantum state  alters it irreversibly. While ``quantum money'' has not turned out to be practical, the basic concept found direct application in cryptography \cite{Singh,Gisin,shanon} or, more precisely in quantum key distribution (QKD)\cite{EBennett,Bennett,E,BennettB}. The two most celebrated QKD protocols, the BB84 \cite{EBennett,Bennett} and E-91\cite{E}, both utilize basic ingredients from ``quantum money'' including that of a qubit.

Based on interaction-free measurements, or quantum interrogation \cite{Elitzur,Kwiat95,Namekata,Jang,dicke}, a QKD protocol was proposed
\cite{guo} which left the door open for a more recent one employing the idea of counterfactuality, meaning no information-carrying qubits need to
travel between Alice and Bob \cite{N} - even though photons in this case can still be found in the transmission channel half the time on average (assuming a 50-50 beam splitter
is used). This protocol was recently realized experimentally \cite{Ren,Brida,P}. One drawback - apart from being nondeterministic - is that, even in the ideal case, only $12.5\%$
of photons used are retained; the rest are discarded.

 The basic idea of interaction-free measurement, central to both counterfactual cryptography and counterfactual computation \cite{Jozsa,Mitchison}, makes use of the fact that the presence of an obstructing object, acting as a measuring device, inside an interferometer setting, destroys interference even if no particle is absorbed by the object. This has the surprising consequence that sometimes the presence of such an object can be inferred without the object directly interacting with any (interrogating) particles.  Using the quantum Zeno effect  \cite{Weinfurter,K,Misra,Peres,Agarwal,H}  (which refers to the fact that repeated measurement of an evolving quantum system can inhibit its evolution, leaving it in its initial state, an effect often paraphrased as ``a watched kettle never boils''), the efficiency of such interaction-free measurements can be dramatically boosted.

Here, we take the logic of counterfactual communication to its natural conclusion. We show how in the ideal limit, using a chained version of the Zeno effect \cite{H}, information can be directly exchanged between Alice and Bob with no physical particles traveling between them, thus achieving direct counterfactual communication.

Our proposed setup is shown in Fig. 1. At Alice's end, it is composed of two  parts. The first part consists of a light source ($S$) that sends a stream of horizontally polarized (H) photons,  detectors (${{D}_{1}}$, ${{D}_{2}}$  and ${{D}_{3}}$), and a polarizing beam splitter $PBS_0$ which only reflects vertically polarized photons V (as do all the $PBS$ in the figure). The second part comprises two tandem Michelson interferometers. It includes two $PBS$s, two switchable polarization rotators ($SRP$s),  two switchable   mirrors ($SM$s) that can be switched on and off by external means, and two normal mirrors ($MR$s). This part of the setup allows the signal photon to have a very large probability of staying at Alice's end. On the other side, at Bob's end, with the help of pockel cell $P{{C}_{B}}$, he can either switch the polarization of the incoming H photon to a  V photon or keep the polarization state H unchanged. The  $PBS_B$ reflects V photons to a detector ${{D}_{4}}$  (effectively blocking the communication channel) and allows H photons to be reflected back by the mirror $MR_B$. Bob can send a stream of logic 0's and 1's by either keeping the polarization state H unchanged (logic 0) or switching it to polarization state V (logic1). Bob's choice of logic 0 and 1 leads to a click at detectors ${{D}_{1}}$ and ${{D}_{2}}$, respectively with almost unit probability and with almost no photon in the transmission channel, thus leading to direct counterfactual communication.

This setup is implementable using current technology. However, before explaining how the setup works, we discuss an equivalent Mach-Zehnder type setup shown in Fig. 2, which helps us to understand the working principle of our protocol. In the Mach-Zehnder setup, $BS$ stands for beam splitter. Initially, a photon is sent by Alice from the left such that the input state (before the top beam splitter) is $\left| \text{10} \right\rangle $. The state transformation at the beam splitters is described by $   \left| \text{10} \right\rangle \to \cos \theta \left| \text{10} \right\rangle +\sin \theta \left| \text{01} \right\rangle$ and $\left| \text{01} \right\rangle \to \cos \theta \left| \text{01} \right\rangle -\sin \theta \left| \text{10} \right\rangle$, where $\cos \theta =\sqrt{R}$ with $R$ being the reflectivity of the $BS$.

At Bob's end, ideal switches  ($SW$) allow Bob to pass the photon (logic 0) or to block it (logic 1).

We now show how to build a direct communication system using the quantum Zeno effect, which refers to the fact that repeated measurement of a gradually evolving quantum state leaves it unchanged.

 Our purpose can be achieved in two steps. In the first step [see Fig.2(a)], we use a large number ($N$) of beam splitters with a very small transmissivity, i.e., $\theta =\pi /2N$. When Bob allows Alice's photon to pass, by switching off all $SW$s at his end, the initial state $\left| \text{10} \right\rangle $ evolves coherently. After $n$ cycles, the state of the photon can be written as
\begin{equation}
    \label{f1}
    \left| \text{10} \right\rangle \to \cos n\theta \left| \text{10} \right\rangle +\sin n\theta \left| \text{01} \right\rangle.
\end{equation}
Thus, at the end of $N$ cycles ($n=N$), the final state is $\left| \text{01} \right\rangle $ and the detector ${{D}_{2}}$ clicks. On the other hand, if Bob blocks the photon by switching on all $SW$s, the photonic state after $n$ cycles is
\begin{equation}
    \label{f2}
    \left| \text{10} \right\rangle \to {{\cos }^{n-1}}\theta (\cos \theta\left| \text{10} \right\rangle + \sin \theta \left| \text{01} \right\rangle) \approx \left| \text{10} \right\rangle,
\end{equation}
where we assumed $N$ to be large and ${{\cos }^{N}}\theta \approx 1$. Here the square of the overall factor (${\cos }^{2(n-1)}\theta$) represents the probability of having the state $\left| \text{10} \right\rangle $ after $n-1$ cycles. In this case the photon is reflected and the detector ${{D}_{1}}$ clicks.

As a result, Bob's blocking causes detector
${{D}_{1}}$ to click, while passing the photon causes detector ${{D}_{2}}$ to click. This means that, in the ideal limit, Alice can read Bob's bit choices with arbitrarily large efficiency. This  is the first step towards direct counterfactual quantum communication.

Although the Mach-Zehnder set-up, shown in Fig. 2(a), enables direct communication, the protocol is only  partially counterfactual. In the case when Bob does not block, the photon's final state $\left| \text{01} \right\rangle $ implies that the photon passes through the transmission channel with unit probability at $N-th$ cycle.

 We now present a protocol that leads not only to direct communication between Alice and Bob but is also fully counterfactual. We use the chained version of the quantum Zeno effect, as shown in  Fig.2(b). The signal photon passes through ``$M$'' big cycles separated by $B{{S}_{M}}$s with ${{\theta }_{M}}=\pi /2M$. For the $m$th cycle ($m\le M$), there are  ``$N$'' beam splitters $B{{S}_{N}}$s with ${{\theta }_{N}}=\pi /2N$.

We assume that initially Alice sends a single photon as shown in Fig. 2(b), where all unused ports are in the vacuum state.
 As a result of beam splitter transformations, now we have three photon states $|i,j,k\rangle$; where $|i\rangle$, $|j\rangle$, and $|k\rangle$ correspond to the photon states at the left-hand side arms of the outer chain, at the left-hand side arms of the inner chain, and at the right-hand side arms of the inner chain,  respectively. By using the results shown in Eqs. \eqref{f1} and \eqref{f2}, it is easy to see that if Bob passes Alice's photon, for the $m-th$ cycle, we have
\begin{equation}
  \label{f3}
    \left| \text{010} \right\rangle \to \cos n{{\theta }_{N}}\left| \text{010} \right\rangle +\sin n{{\theta }_{N}}\left| \text{001} \right\rangle \overset{n=N}{\mathop{\to }}\,\left| \text{001} \right\rangle.
\end{equation}

The initial state of the total system is $\left| \text{100} \right\rangle $. We can see the evolution by including results from Eqs. \eqref{f1} and \eqref{f2}.

First we consider the case when Bob does not block at any stage (logic 0). After the $m-th$ cycle, the resulting photon state is
\begin{equation}
    \begin{split}
    \left| \text{100} \right\rangle &\to {{\cos }^{m-1}}{{\theta }_{M}}({\cos }{{\theta }_{M}}\left| \text{100} \right\rangle +\sin {{\theta }_{M}}\left| \text{010}) \right\rangle\\ &\overset{m=M}{\mathop{\to }}\,\left| \text{100} \right\rangle.
    \end{split}
 \end{equation}
It is clear that after M big cycles and N small cycles detector ${{D}_{1}}$  clicks. A click at the detector ${{D}_{1}}$ ensures counterfactuality as any photon in the transmission channel would lead to a click at one of the detectors ${{D}_{3}}$ [see Eq.  \eqref{f1}]. The probability of click at ${{D}_{1}}$ is obtained by collecting all the contributions that are reflected from all the beam splitters $BS_m$'s and is given by $P_1= {\cos}^{2M}{\theta_M}$.

On the other hand, if Bob blocks throughout (logic 1), we have (for the $m-th$ cycle)
	\begin{equation}
    \begin{split}
    \label{f4}
    \left| \text{010} \right\rangle &\to {{\cos }^{n-1}}{{\theta }_{N}}({\cos }{{\theta }_{N}}\left| \text{010} \right\rangle +\sin {{\theta }_{N}}\left| \text{001} )\right\rangle\\ &\overset{n=N}{\mathop{\to }}\,\left| \text{010} \right\rangle,
    \end{split}
    \end{equation}
where we assume $N>>1$. After the $m-th$ cycle, the photon state is
	\begin{equation}
    \left| \text{100} \right\rangle \to \cos m{{\theta }_{M}}\left| \text{100} \right\rangle +\sin m{{\theta }_{M}}\left| \text{010} \right\rangle \overset{m=M}{\mathop{\to }}\,\left| \text{010} \right\rangle.
    \end{equation}
Thus after $M$ big cycles and $N$ small cycles, detector ${{D}_{2}}$ clicks. Again counterfactuality is ensured by a click at ${{D}_{2}}$  as any photon in the transmission channel would be absorbed by the blocking device and would not be available for detection at ${{D}_{2}}$.  The probability of click at the detector ${{D}_{2}}$ is given by $P_2= |y_{\{M,0\}}|^2$, where $y_{ \{M,0\} }$ can be obtained from the  recursion relations $x_{m+1} = a_M x_{m} - b_M y_{\{m,N\}}$, $y_{\{m+1,0\}}=b_M x_{m}+a_M y_{\{m,N\}}$, $y_{\{m,n\}}=a_N y_{\{m,n-1\}}-b_N z_{\{m,n-1\}}$ and $z_{\{m,n\}} = c (b_N y_{\{m,n-1\}} + a_N z_{\{m,n-1\}})$ where  $a_{N(M)} ={\cos} \theta_{N(M)}$, $b_{N(M)}={\sin} \theta_{N(M)}$, and $c = 0$ with $x_1 = a_M$, $y_{\{1,0\}} = b_M$ and $z_{\{m,0\}}=0$. Obviously, if $c=1$, we can get the probability $D_1$ clicking ($P_1=|x_M|^2$) for the case Bob encoding ``0''.

We emphasize that for $D_1$ or $D_2$ clicking, no photon could have passed through the transmission channel, since the presence of any photon in the channel would have led to detection events at $D_3$ (for Bob does not block) or at Bob's blocking device (for Bob blocks).

In Figs. (3a) and (3b), we have plotted the probabilities $P_1$ and $P_2$ by using the above recursion relations. It is clearly seen that $P_1$ is above 0.90 for $M>25$ and is independent of $N$; however, a value of $P_2$  above 0.90 requires a much larger value of $N$. Our numerical estimates indicate ($P_1=0.906$, $P_2=0.912$)for ($M=25$, $N=320$);  ($P_1=0.952$, $P_2=0.953$) for ($M=50$, $N=1250$), and  ($P_1=0.984$, $P_2=0.982$) for ($M=150$, $N=10000$). This shows that perfect conterfactuality is possible, albeit for large values of $M$ and $N$.

This may be complicated for the Mach-Zehnder setup discussed so far. However a Michelson interferometer-based implementation offers significant practical advantages. Thus, after elucidating the essential features of our direct counterfactual quantum communication protocol, we revert to a discussion of the Michelson-type configuration  shown in Fig. 1. This allows a better practical realization of the protocol, with a massive saving of resources.

Here, the function of $BS$ is replaced by the combination of $PBS$ and $SPR$. Assume the state of an H photon is $\left| H \right\rangle $, and the state of a V photon is $\left| V \right\rangle $. Then, each time the photon passes through one $SPR$, the polarization evolves as follows $ \left| H \right\rangle \to \cos {{\beta }_{i}}\left| H \right\rangle +\sin {{\beta }_{i}}\left| V \right\rangle$ and $ \left| V \right\rangle \to \cos {{\beta }_{i}}\left| V \right\rangle -\sin {{\beta }_{i}}\left| H \right\rangle$, where $\beta$ represents the rotation angle with the subscript $i=1,2$ corresponding to different $SPR$s. The mirror $S{{M}_{1(2)}}$ is switched off initially to allow the photon to be transmitted but it remains on during $M(N)$ cycles and is turned off again after $M(N)$ cycles are completed. The initial photon emitted by the light source is $\left| H \right\rangle $. Since the signal photon passes through $SM$s twice each cycle, we set ${{\beta}_{1(2)}}=\pi/4M(N)$. It is not difficult to see that if Bob blocks the photon, detector ${{D}_{2}}$ clicks. Also, if Bob passes the photon, detector ${{D}_{1}}$ clicks.

Next we consider the effect of imperfections of the system and noise in the transmission channel on the performance of counterfactual communication. There are two kinds of imperfections: The first one only affects the efficiency of communication, but does not cause measurement errors. Imperfection coming from the sensitivity of detectors $D_1$ and $D_2$ is an example of this. If the sensitivity of these detectors is $\eta$, then the efficiency of communication also reduces to $\eta$. However, the second kind of imperfection, which mainly comes from the switchable polarization rotators ($SPR$s), results in measurement errors. During each cycle, $SPR$s should rotate the signal photon with a certain angle, but in practical situations there can be a slight error in the angle. Let us suppose that the error for the $SPR$ in the inner cycle is  $\Delta {{\theta }_{N}}\text{=}{{s}_{N}}({{\theta }_{N}}/N)$; namely, the photon state is rotated with an additional angle $s_N \theta_N$ after $N$ cycles.  The corresponding coefficient for the error of the $SPR$ in the outer cycle is $s_M$. We can estimate their influence numerically by replacing $\theta_{N(M)}$ with $\theta_{N(M)}+\Delta \theta_{N(M)}$ for fixed $N$ and $M$ in the recursion relations given above. In Fig. 4(a), we plot the detector successful clicking rates for different values of $s$ (setting $s=s_N=s_M$). It is clear that the performance is still good if the factor $s$ is less than two. In Fig. 4, we also show the error rate associated with the wrong clicking of $D_1$ and $D_2$ by using the concept of mutual information $I(X,Y)$. We consider a communication process in which Bob sends messages composed of logic 0 and 1 with equal probabilities. Let the ensemble `X' represent the detector `x' clicking, with $x=D_1,D_2,D_3,D_4$. Also the events $y\in \text{Y}$ correspond to the clicking of detectors $D_1$ and $D_2$ giving wrong information; i.e., $y=D_1$ represents Bob sending ``1'' (Alice's $D_1$ incorrectly clicking instead of $D_2$) and $y=D_2$ represents Bob sending ``0'' (Alice's $D_2$ incorrectly clicking instead of $D_1$). Then, mutual information can be defined as

\begin{equation}
\begin{split}
\label{f5}
I(X,Y)&=\sum\limits_{x,y}{P(x,y)\log \frac{P(x,y)}{P(x)P(y)}}  \\
&=-\sum\limits_{i=1,2}{P(y={{D}_{i}})\log P(x={{D}_{i}})}
\end{split}
\end{equation}
It is easy to see that if the error rate is zero, the mutual information is zero.

Another source of noise results when the transmission channel is blocked by an object other than Bob's. We can define the noise rate as $B$. This represents the probability of any object other than Bob's blocking the channel. It is easy to see if Bob chooses to block his path, the result at Alice's end does not change. For the case when Bob allows the photon component to be reflected, the result again does not change appreciably if there is blocking only in one cycle. However, the noise may cause a problem if blocking takes place in multiple cycles. In Fig. 4(b), we plot the probability of $D_1$ clicking for different values of $B$ as well as the mutual information. To simulate the noise, we create random numbers between 0 and 1 each time the photon component passes through the transmission channel. If the number is less than $B$, we take the transmission channel to be blocked (set $c=0$ for that cycle, otherwise $c=1$). The figure shows that the blocking rate $B$ should be suppressed under 0.2\%. A higher loss may adversely affect our protocol.

We also note that the time control of switchable mirrors ($SM$s) is also very important. Suppose the distance between Alice and Bob is $L$. The control time of these switchable mirrors should be less than $2L/c_0$ ($c_0$ being the light speed).

In summary, we strongly challenge the longstanding assumption that information transfer requires physical particles to travel between sender and receiver by proposing a direct quantum communication protocol whereby, in the ideal asymptotic limit, no photons  pass through the transmission channel, thus achieving complete counterfactuality. In so doing we highlight the essential difference between classical and quantum information.

\begin{acknowledgments}
This research is supported by a grant from King Abdulaziz City for Science and Technology (KACST) and an NPRP grant (4-520-1-0830) from Qatar National Research Fund (QNRF).
\end{acknowledgments}


   \begin{figure}[t]
    \begin{center}
   \begin{tabular}{c}
    \includegraphics[height=6.25cm,width=0.99\columnwidth]{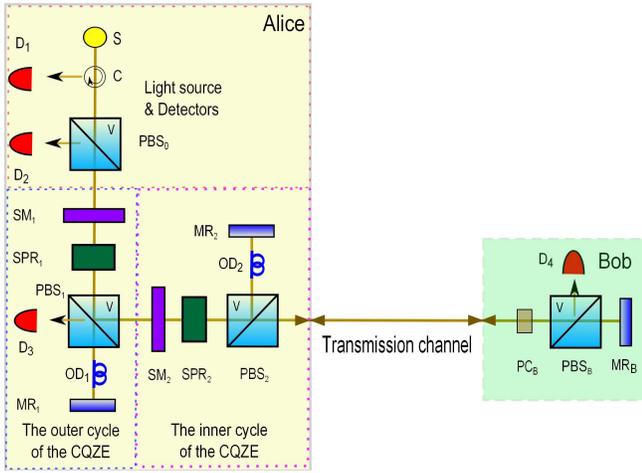}
    \end{tabular}
    \end{center}
\begin{widetext}
   \caption[2Dpattern]
   { \label{fig:fig1}
 (Color online)
In the figure, $S$ stands for the light source, $C$ is the optical circulator, ${{D}_{1}}$, ${{D}_{2}}$, ${{D}_{3}}$ and ${{D}_{4}}$ are photon detectors, $PBS$ stands for polarizing beam splitter which only reflects vertically polarized photons (V), SPR stands for switchable polarization rotator, $PC$ stands for Pockels cell that determines the polarization state of the transmitted photons, $SM$ stands for switchable mirror, $MR$ stands for a normal mirror and $OD$ stands for optical delay. Only horizontally polarized photons (H) will be sent into the tandem Michelson interferometers. The two optical paths $S{{M}_{1}}\rightarrow M{{R}_{1}}$ and $S{{M}_{1}}\rightarrow M{{R}_{B}}$ for the first Michelson interferometer correspond to the outer cycle of the chained quantum Zeno effect CQZE ($M$ cycles) for Mach-Zehnder setup, while the paths $S{{M}_{2}}\rightarrow M{{R}_{2}}$ and $S{{M}_{2}}\rightarrow M{{R}_{B}}$ for the second Michelson interferometer, correspond to the inner cycle of the CQZE ($N$ cycles). The mirror $S{{M}_{1(2)}}$ is switched off initially to allow the photon to be transmitted but it then remains on for $M(N)$ cycles, and is turned off again after $M(N)$ cycles are completed. Here $SP{{R}_{1(2)}}$ rotates the polarization by a small angle ${{\beta }_{M(N)}}=\pi /4M(N)$ (for each cycle, the photon passes $SPR$ twice), i.e., $\left| H \right\rangle $ evolves to $\cos {{\beta }_{M(N)}}\left| H \right\rangle +\sin {{\beta }_{M(N)}}\left| V \right\rangle $ and $\left| V \right\rangle $ evolves to $\cos {{\beta }_{M(N)}}\left| V \right\rangle -\sin {{\beta }_{M(N)}}\left| H \right\rangle $. $O{{D}_{1}}$ and $O{{D}_{2}}$ guarantee that optical distances of different paths of same interferometer exactly match. At Bob's end, Bob passes an H photon by turning off his $PC$ reflecting it back, and he blocks an H photon by turning on his $PC$, changing the photon's polarization to V. We  emphasize that the chance of Alice's photon leaking into the transmission channel is almost zero for large enough $M$ and $N$.
}
\end{widetext}
  \end{figure}

 \begin{figure}[t]
    \begin{center}
   \begin{tabular}{c}
    \includegraphics[height=10.1cm,width=0.99\columnwidth]{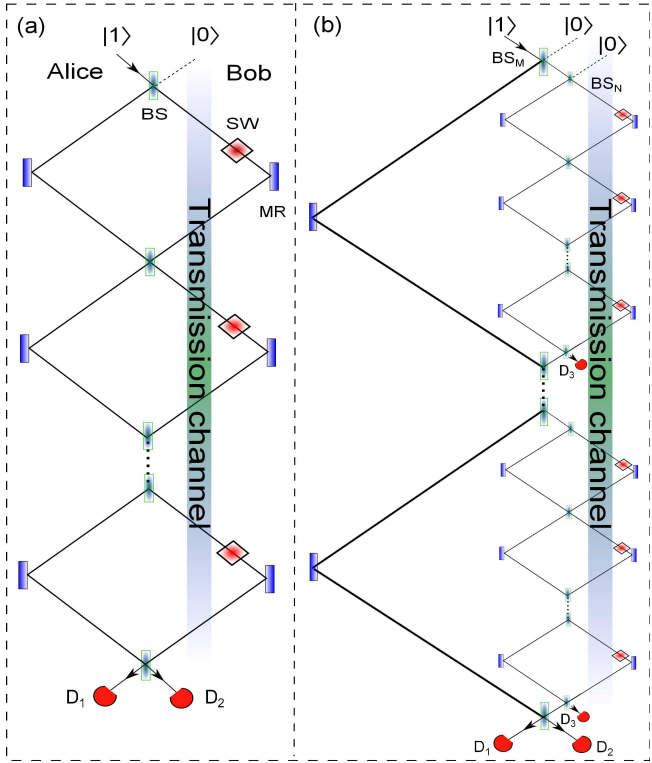}
    \end{tabular}
    \end{center}
   \caption[2Dpattern]
   { \label{fig:fig2}
 (Color online) Here $BS$ stands for beam splitter and $SW$ stands for ideal switches. In the transmission channel, the photon is accessible to Eve. (a) The $BS$s have large reflectivity, $R={{\cos }^{\text{2}}}\theta \text{=}{{\cos }^{\text{2}}}(\pi /2N)$ with $N$ being the total number of beam splitters. (b) By using a chained version of the setup shown in (a), we can achieve direct counterfactual quantum communication. Two kinds of $BS$s are used. One is $B{{S}_{M}}$ for $M$ big cycles. The other is $B{{S}_{N}}$ for $N$ small cycles within each $M$ cycle. There are a total of $M\times N$ cycles for one signal. As discussed in the text, the probability of finding a signal photon in the transmission channel is nearly zero. Clicks at ${{D}_{1}}$ or ${{D}_{2}}$ reveal to Alice Bob's bit choices.}
  \end{figure}

\begin{figure}[t]
    \begin{center}
   \begin{tabular}{c}
    \includegraphics[height=4.1cm,width=0.99\columnwidth]{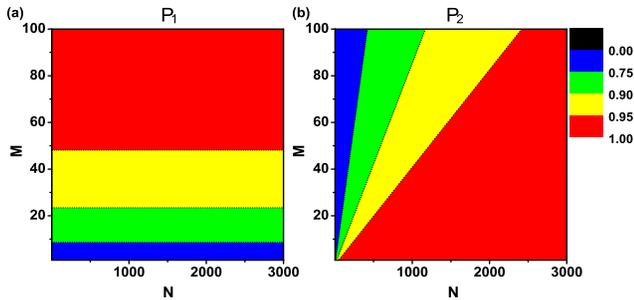}
    \end{tabular}
    \end{center}
   \caption[2Dpattern]
   { \label{fig:fig3}
  (Color online) ${{P}_{1}}$ and ${{P}_{2}}$, which are the probabilities of ${{D}_{1}}$ and ${{D}_{2}}$ clicking, respectively, are plotted against different number of cycles $M$ and $N$ for (a) Bob unblocking Alice's photon and (b) Bob obstructing Alice's photon.}
\end{figure}

\begin{figure}[t]
    \begin{center}
   \begin{tabular}{c}
    \includegraphics[height=4.0cm,width=0.99\columnwidth]{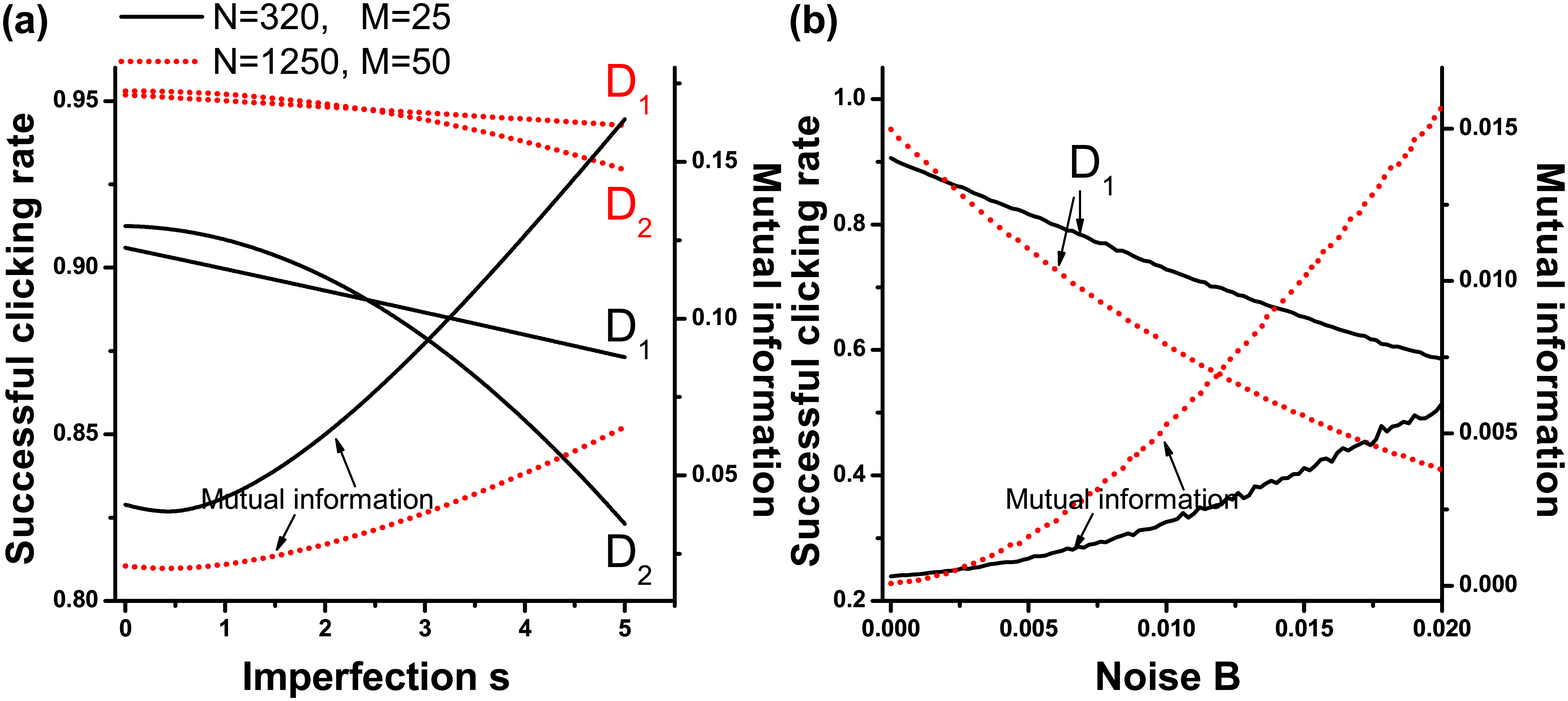}
    \end{tabular}
    \end{center}
   \caption[2Dpattern]
   { \label{fig:fig4}
  (Color online) (a)The variation of the rate of successful clicking plotted for ${{D}_{1}}$ and ${{D}_{2}}$ as a function of $s$, where $s$ describes the imperfection of the switchable polarization rotators. Also plotted is the mutual information describing the error rate of $D_1$ and $D_2$ as a function of $s$. (b) The rate of successful clicking of ${{D}_{1}}$ and the corresponding mutual information both plotted as a function of noise $B$, defined as the probability of any object other than Bob's blocking the transmission channel. The red dotted curves are plotted for the case $M$=50, $N$=1250. The black solid curves are plotted for the case $M$=25, $N$=320.}
\end{figure}


\end{document}